# Spatially modulated heavy-fermion superconductivity in CeIrIn$_5$

Maja D. Bachmann[1,2,†], G. M. Ferguson[3,†], Florian Theuss[3], Tobias Meng[4], Carsten Putzke[1], Toni Helm[1], K.R. Shirer[1], You-Sheng Li[1,2], K.A. Modic[1], Michael Nicklas[1], Markus König[1], D. Low[3], Sayak Ghosh[3], Andrew P. Mackenzie[1,2], Frank Arnold[1], Elena Hassinger[1], Ross D. McDonald[5], Laurel E. Winter[5], Eric D. Bauer[5], Filip Ronning[5], B.J. Ramshaw[3], Katja C. Nowack[3,6,*,‡] and Philip J.W. Moll[1,*,‡]

[1]Max-Planck-Institute for Chemical Physics of Solids, Dresden, D-01187 Germany
[2]School of Physics and Astronomy, University of St. Andrews, St. Andrews KY16 9SS, UK
[3]Laboratory of Atomic and Solid State Physics, Cornell University, Ithaca, NY 14853, USA
[4]Institute for Theoretical Physics, Technical University Dresden, D-01062 Dresden, Germany
[5]Los Alamos National Laboratory, Los Alamos, NM 87545, USA
[6]Kavli Institute at Cornell, Ithaca, New York 14853, USA
* To whom correspondence should be addressed: philip.moll@epfl.ch, kcn34@cornell.edu
[†]These authors contributed equally to this work.
[‡]These authors contributed equally to this work.

**The ability to spatially modulate the electronic properties of solids has led to landmark discoveries in condensed matter physics as well as new electronic applications. Although crystals of strongly correlated metals exhibit a diverse set of electronic ground states, few approaches to spatially modulating their properties exist. Here we demonstrate spatial control over the superconducting state in mesoscale samples of the canonical heavy-fermion superconductor CeIrIn$_5$. We use a focused ion beam (FIB) to pattern crystals on the microscale, which tailors the strain induced by differential thermal contraction into specific areas of the device. The resulting non-uniform strain fields induce complex patterns of superconductivity due to the strong dependence of the transition temperature on the strength and direction of strain. Electrical transport and magnetic imaging of devices with different geometry show that the obtained spatial modulation of superconductivity agrees with predictions based on finite element simulations. These results present a generic approach to manipulating electronic order on micrometer length scales in strongly correlated matter.**

Heavy fermion materials exhibit a rich competition between metallic, superconducting, and magnetically ordered ground states. The ability to locally control electronic properties within these materials would enable the design of new correlated states both for fundamental research and for applications. Alternative approaches to achieve spatially modulated correlations involve modulating the chemical composition, for example in epitaxially grown heterostructures (1). Here we demonstrate mesoscopic control over the superconducting order in stoichiometric and ultra-clean CeIrIn$_5$ by inducing a tailored strain field in microstructured single crystal devices. Our experimental approach is based on the interplay between strain induced by differential thermal contraction between the sample and the substrate, and sub-μm control over the shape of the sample.

The heavy-fermion CeIrIn$_5$ provides ideal material parameters to realize the spatial modulation of a correlated state (Sommerfeld coefficient $\gamma \sim 720$ mJ mol$^{-1}$ K$^{-2}$ (2), effective mass $m^* \sim 30 m_e$ (3)). The superconductivity is highly sensitive to strain, due to the strong dependence of Ce-4f hybridization on the Ce-Ce interatomic distance. Straining the sample along the crystallographic a-direction increases the bulk superconducting transition temperature ($T_{c0}$ = 400 mK) by 56mK/kbar, while compression along c decreases it by -66mK/kbar (4, 5). Although uniaxial strain strongly alters $T_c$, the almost equal but opposite effects of a- and c-direction strain lead to an overall weak change of $T_c$ under hydrostatic pressure (10mK/kbar ~ 2.5% $T_c$ / kbar) (6). The anisotropic response of $T_c$ to strain makes spatial modulation of superconductivity possible in a single crystal of CeIrIn$_5$: micrometer scale control over the strain field in the crystal leads to micrometer scale variation in $T_c$ without chemical inhomogeneities. Single-material Josephson junctions may be realized by local suppression of the order parameter in selected areas of a crystal, and clean interfaces between magnetic and superconducting order within a single crystal would facilitate new investigations of quantum critical phenomena.

## Non-uniform superconductivity within a single crystal

A simple rectangular lamella serves as an instructive testing ground for this concept (Fig.1). The lamella (150 x 30 x 2 μm$^3$) was carved from a macroscopic crystal using Focused Ion Beam machining (for details on the fabrication, see (7–9)) and joined to the sapphire substrate by a thin layer of epoxy (~ a few hundred nanometers). The crystallographic c-direction is aligned with the short and the a-direction with the long side of the lamella. Using (0001)-cut sapphire ensures an isotropic thermal contraction of the substrate. While sapphire is known for its low thermal contraction, CeIrIn$_5$ contracts strongly upon cooling as is typical of many Ce-compounds (10). As a result, the sample is under tensile strain at low temperature.

To study the superconducting transition in the lamella, we use scanning superconducting quantum interference device (SQUID) microscopy (SSM) to image the diamagnetic response of the sample with micrometer scale resolution. To detect superconductivity, we apply a local magnetic field using a ~6μm field coil integrated on the SQUID chip while monitoring the local magnetic susceptibility with a ~1.5μm

SQUID pickup loop (see (9) for details). Superconducting regions of the sample exhibit a strong diamagnetic response, allowing us to distinguish them from metallic and insulating regions.

Susceptibility images as a function of temperature (Fig.1) show that superconductivity first emerges at the short edges while the majority of the lamella remains metallic. As the temperature is lowered, larger fractions of the sample become superconducting, leading to the growth of triangular superconducting patches protruding into the slab, which eventually join in the center. At even lower temperatures, the order parameter remains suppressed on the long edge. The observation of superconductivity at the edge of the sample before the interior is unexpected; for a thin superconducting slab cooled in the earth's field, the demagnetization factor at the sample edges favors the appearance of superconductivity in the center of the sample first (11).

Before analyzing the spatial pattern in the images, we estimate if strain due to differential thermal contraction combined with the strain sensitivity of $T_c$ in $CeIrIn_5$ can account for the observed variations in $T_c$ of several 100 mK. When cooled to cryogenic temperatures, $CeIrIn_5$ and sapphire contract by ~0.3% and ~0.08% respectively. Given this mismatch, we expect strain on the order of 0.1% to exist within the $CeIrIn_5$ crystal at low temperature. Previous studies on bulk single crystals (4, 5) show that ~0.1 GPa of uniaxial pressure changes $T_c$ by 100 mK. Using a typical elastic modulus of 150 GPa, we estimate that compressive strains of ~0.1% are achieved in these experiments. Both uniaxial pressure studies and the image series presented here are consistent with ~0.1% strain generating ~100 mK variation in $T_c$.

To understand the geometry in the images, we perform finite element method simulations of the strain field in the device caused by the difference in the thermal contraction between $CeIrIn_5$ and sapphire (9). We then compute the local transition temperature from the strain field using,

$$T_c = T_{c0} + \frac{\delta T_c}{\delta \varepsilon_a}(\varepsilon_a + \varepsilon_b) + \frac{\delta T_c}{\delta \varepsilon_c}\varepsilon_c,$$

for each point on the grid resulting in a spatial map (Fig. 1 C). Here $\varepsilon_i$ with $i = a, b, c$ are the diagonal elements of the strain tensor along the corresponding crystallographic directions. We estimate $\frac{\delta T_c}{\delta \varepsilon_a} = -57\ K$, $\frac{\delta T_c}{\delta \varepsilon_c} = 66\ K$ from reported bulk measurements of $T_c$ as a function of uniaxial pressure and our measured elastic moduli (4, 5, 9). From this map we generate binary images marking superconducting and metallic regions at each temperature that can be directly compared to our susceptibility images (Fig. 1 D-F). We find detailed agreement between the spatial patterns of superconducting regions observed in the SSM images at different temperatures and the pattern predicted by the simulations, indicating that our modeling captures the physical origin of the complex superconducting patches. Given the uncertainties in the thickness and the arrangement of the adhesive layer bonding the sample and substrate, the quantitative agreement between the experimental data and the ab-initio calculation without fitting parameters is remarkable.

**Induced strain field depends on geometry of FIB defined features**

Next we show that the induced strain field and the shape of the superconducting regions can be tailored by FIB defining the $CeIrIn_5$ microstructure. In (Fig.2) we show two $CeIrIn_5$ lamellas with additional trenches cut through the crystal down to the substrate. In both devices the trenches define a square in the (a,c)-plane. In device 1 the square is anchored by 4 constrictions in each corner (Fig. 2A). In device 2, the constrictions connect to the center of each side of the square (Fig. 2F). Under biaxial tension, each of the contact pads is pulled outwards, subjecting the square to non-uniform strain.

Device 1 is designed to measure anisotropic resistances in the plane by passing current through any pair of neighboring contacts, while measuring voltage across the remaining pair (*12*, *13*). In this device, the simulated $T_c$ map predicts a pattern similar to the unstructured bar, with the first superconducting regions developing on the edges aligned with the c-direction as the device is cooled (Fig. 2 F). These regions extend towards the center upon lowering the temperature, eventually connecting in the middle of the device. These patterns are clearly observed in SSM images (see Fig. 2 B-D) and lead to three distinct regimes for transport through device 1: a) the normal state in which all contacts are separated by metallic regions, b) a state in which only the contact pairs along the c-direction are connected by superconducting regions, c) a state in which all contacts are connected by a single superconducting region. As a result, if current is sourced between contacts along the c-direction (1 and 2) a transition to zero voltage signaling superconductivity at a relatively high temperature $T_c^*$ is observed (Fig. 2 E, red trace). On the other hand, for currents sourced between contacts along the a-direction (2 and 4), (Fig. 2 E, blue trace) we observe a sharp upturn in voltage along a as $R_c$ goes to zero. The elongated superconducting regions identified in Fig. 2D cause the current from contacts 2 to 4 to distribute evenly over the width of the device, leading to a larger current to flow at the voltage probes (1 and 3). Eventually, a second transition to a zero-resistance state is observed when all terminals are connected by a single superconducting region.

Superconductivity emerges in a strikingly different spatial pattern in device 2 (Fig. 2 H-K). In contrast to device 1, the edges in device 2 parallel to the a-axis superconduct well before the central region of the device. As with device 1, we find that simulations of the strain profile in the device reproduce the structure of the superconducting transition in detail. The contrast between the two image series indicates that the structure in the images is determined by the interplay between the intrinsic strain sensitivity of the material and the strain field imposed by the FIB defined features.

We observe a pronounced suppression and enhancement of $T_c$ in the a- and c-aligned constrictions respectively. In essence, the strain in the constrictions is enhanced due to the contact pads on one side and the square on the other side exerting forces on the constrictions which point outwards. This provides a route to design devices that exhibit a strong modulation of $T_c$ and to generate confined regions of suppressed superconductivity.

### Large tunability of $T_c$ in small structures

Smaller devices exhibit an even more pronounced dependence of the transport $T_c$ on their geometry. Device 3 features three series-connected straight beams with dimensions (22x1.7x8 $\mu m^3$) with two beams aligned with the c- and one with the a-direction. These fine structures cannot be resolved in detail with the SSM. In transport, we find that the transition temperature for the c-aligned beam $T_c^c \sim 700 mK$ is higher than the bulk $T_c$, while the transition temperatures for the a-aligned beams $T_c^a \sim 200 mK$ are significantly lower than the bulk $T_c$. Additional small structures show equally or even more dramatic variations of $T_c$ (9). Therefore, a modulation of the transition temperature by more than a factor of 4 within a single crystal can be realized within our fabrication approach.

### High critical currents

To exploit strain tuning of the superconducting order for future device-based experiments, a key question concerns the robustness of the induced superconducting state. Figure 3 shows characterization of device 3 at high applied currents to estimate the critical current of each beam. The current is increased until an observable voltage signals the breakdown of the zero-resistance state. To minimize self-heating, rectangular current pulses of duration of 83$\mu$s are applied to the sample with a cooldown time between pulses of 100ms. Because the critical current decreases monotonically with increasing temperature, the obtained values represent a lower bound of their magnitude in the absence of heating. Figure 3A shows a typical current-voltage characteristic of the c-aligned beam at 500mK, well above the bulk $T_c$. A robust zero resistance state is detected up to a critical current density of about 12.5 kA cm$^{-2}$. Upon cooling, this quantity can be extrapolated to $j_c^c$ (0K) ~ 18 kA

cm$^{-2}$ (Fig. 3B). This high critical current is typical for bulk heavy fermion superconductors (UPt$_3$ ~3.8 kA cm$^{-2}$ ref.(*14*); URu$_2$Si$_2$ ~24 ref.(*14*); CeCu$_2$Si$_2$ ~1-5 ref.(*15*)), and thereby strongly supports the scenario of robust, bulk-like superconductivity over the formation of sparse superconducting filaments under strain.

## 1K phase in CeIrIn$_5$

An open puzzle in CeIrIn$_5$ is the origin of the commonly observed discrepancy between thermodynamic and resistive measurements of T$_c$ in CeIrIn$_5$. Bulk crystals show a transition to a zero resistance state well above T$_c$, starting as high as T$_c$*~1.2K (*16–18*) dependent on the details of the sample. Our experiments suggest that strain is responsible for both the 1K phase and the modulation in T$_c$ observed in the microstructured devices.

This is further corroborated by comparing the magnetic field dependence of the strain-modified superconductivity in device 3 and the 1K-phase. For the 1K phase the upper critical field estimated from transport collapses onto the same curve as the critical field observed in thermodynamic measurements for the bulk by simple scaling (*2*). We estimate the critical fields $H_{c2}^a$ and $H_{c2}^c$ using transport on different parts of device 3, corresponding to regions of different strain-tuned T$_c$. The temperature and angle dependences of both critical fields collapse onto the same curve when scaled by the respective values of T$_c$ and H$_{c2}$ (Fig. 4B and C). While these two are clearly different in value, both critical fields are characterized by a similar anisotropy, $H_{c2}^a / H_{c2}^c$ ~ $H_{c2}^{*,a} / H_{c2}^{*,c}$ ~ 2, which is consistent with previous reports (*19*). These observations indicate that our microstructure experiments and the measurements of the 1K-phase on macroscopic samples probe the same underlying physics of T$_c$-enhancement due to strain.

Local strain fields around impurities have been proposed as a mechanism to induce this anomalous 1K-phase (*19–21*), similar to the well-studied "3K-phase" of Sr$_2$RuO$_4$(*22*). However, microscopic analyses have shown the absence of large scale bulk inclusions or crystal defects in the high quality crystals of CeIrIn$_5$ (*20*), rendering this scenario unlikely. We propose that the origin of the 1K phase regularly observed in single crystals arises from surface stresses due to handling of the crystal, and in particular from wiresawing. Macroscopic samples prepared by wiresaw cutting for this study exhibit zero resistance around 1.2K. Resistive signatures of this 1K-phase are completely removed by short surface etching in HCl, after which the resistive transition coincides with thermodynamic probes at T$_c$~400mK (*9*). This demonstrates the absence of defect-strained superconducting patches in good single crystals of CeIrIn$_5$ and resolves the issue as an effect of surface strain.

## Discussion and outlook

We demonstrate a route to spatially modulate superconductivity within a homogeneous electronic system induced by strain. Different mechanisms can result in the spatial gradients of $T_c$, for example by modulating the chemical potential across the sample. Chemical doping landscapes have been achieved in a controlled fashion by gradient-doping in molecular-beam epitaxy techniques(23). Here, controlling the local charge carrier density $n(E_F)$ modifies the local transition temperature. In the simplest case of conventional superconductors, this change can be captured by BCS theory as $T_c \propto e^{-\frac{1}{\lambda n(E_F)}}$, where $\lambda$ denotes the coupling strength, yet $n(E_F)$ also tunes $T_c$ in unconventional superconductors. In the stoichiometric $CeIrIn_5$ microdevices, the charge carrier density is uniform as evidenced by the observation of unperturbed quantum oscillations in the device 3 (Fig. 5). The quantum oscillations quantitatively match the angle dependence of previously reported de Haas-van Alphen oscillations(3) measured on macroscopic crystals and indicate that the Fermi surface shape remains unchanged by the strain field. In particular, the large, heavy orbits and their fine structure are readily observed which is usually very difficult in transport. This is incompatible with the presence of strong charge carrier density changes across the sample, which would lead to spatial variations in the Fermi surface cross-sections and subsequently strongly suppress quantum oscillations by phase smearing.

At the same time, the strain fields in this device are strong enough to modulate $T_c$ by almost a factor of 4, from 200mK to 780mK. This suggest that the strain field spatially modulates the degree of 4f-hybridization across the device, and thereby $T_c$. This at first surprising mechanism is compatible with experimental observations in the related compound $CeRhIn_5$. Here, hydrostatic pressure suppresses anti-ferromagnetism and eventually induces superconductivity. Despite the significant changes in the 4f-magnetism and the spin fluctuation spectrum, the quantum oscillation frequencies remain completely unchanged in the entire pressure range up to the quantum critical point(24). The 4f-hybridization, however, increases as evidenced by the growing quasiparticle effective mass upon applying pressure. This strongly supports the concept that modifying the hybridization with the 4f electrons can leave the overall Fermi surface volume unchanged. Here we propose that the same microscopic physics underlies the spatial modulation of $T_c$ in $CeIrIn_5$ microstructures.

In general, strongly correlated materials exhibit a pronounced sensitivity to perturbations, owing to the small energy scales defining their physics. The strain accessible using our fabrication approach is sufficient to substantially alter the electronic properties of these materials (27), without introducing chemical disorder.

Unlike more conventional approaches to strain-tuning, the FIB provides micrometer scale control over both the direction and magnitude of the induced strain field. We expect that the approach demonstrated here will enable spatial control of a wide range of broken symmetry states in strongly correlated systems. In particular, we envision clean interfaces between regions of the sample with different electronic order generated by a spatially modulated strain field (CeRhIn5, (24)). This fabrication is compatible with an array of scanning probes, leading the way to a range of new experiments on strongly correlated matter. In summary, our approach of design and fabrication of complex superconducting structures within a clean single crystal is of interest for both fundamental and applied research. The results show a surprising new angle to fabricate S/N/S Josephson junctions within a single crystal. On a technological side, it remains to be seen if heavy fermion materials will have applications in quantum technologies. As a mature materials class of unconventional superconductors at highest quality, these d-wave superconductors offer new approaches to spatially manipulate quantum phases. Strain engineering may offer alternative ways to generate superconducting circuitry in a homogeneous metallic device without any physical junctions in the future.

**Acknowledgments**: We are very grateful for the interesting and stimulating discussions with S. Kivelson, J. Zaanen, J. Tranquada, M. Vojta, P. Fulde, J. Thomson, S. Wirth, M. Sigrist, C. Geibel, H. von Löhneysen. We thank N. Nandi for the low temperature measurements.

**Funding**: Work at the Max Planck Institute of Chemical Physics of Solids was supported by the Max-Planck-Society and funded by the Deutsche Forschungsgemeinschaft (DFG, German Research Foundation) – MO 3077/1-1. Work at Cornell University was supported by the Department of Energy, Office of Basic Energy Sciences, Division of Materials Sciences and Engineering, under Award DE-SC0015947 (scanning SQUID imaging, implementation of mK microscope) and by the Cornell Center of Materials Research with funding from the NSF MRSEC program under Award DMR-1719875 (SQUID and microscope design). TM is supported by the Deutsche Forschungsgemeinschaft through SFB 1143 and the Emmy Noether-Programme via grant ME 4844/1-1. M.D.B. acknowledges studentship funding from EPSRC under grant no. EP/I007002/1. P.J.W.M. was supported by the European Research Council (ERC) under the European Union's Horizon 2020 research and innovation programme (GA N° 715730). Work at Los Alamos National Laboratory was performed under the auspices of the US Department of Energy, Office of Basic Energy Sciences, Division of Materials Sciences and Engineering.

**Author contributions:** M.D.B, K.A.M., M.K., C.P., P.J.W.M. fabricated the microstructures, M.D.B., T.H., K.R.S. K.A.M., Y.-S.L., M.N., C.P., P.J.W.M. performed the transport measurements and G.M.F. performed the scanning SQUID imaging with support from D.H.L. and K.C.N.. F.T. and G.M.F. performed the finite element simulations with input from K.C.N. and B.J.R.. The crystals were grown by E.D.B and F.R. T.M. contributed the theoretical treatment of a 1D pairing. F.A., E.H., M.D.B. performed the magneto-transport measurements in the dilution refrigerator. R.D.M. and L.W. measured simultaneous transport in liquid 3He. All authors were involved in the design of the experiment and writing of the manuscript.

**Data availability:** All data needed to evaluate the conclusions in the paper are present in the paper. Additional data related to this paper may be requested from the authors. All data underpinning this publication can be accessed in comprehensible ASCII format at the pure research information system of the University of St Andrews at www.st-andrews.ac.uk/staff/research/pure.

**Supplementary information:** Additional data supporting this manuscript is available online. It contains subsections: Single-crystal growth; FIB fabrication; Crystalline quality of the microstructures; Reproducibility; Susceptibility images; Finite element method simulations; Measurements of $CeIrIn_5$ elastic moduli.

**Preprint:** A preprint including supplementary material can be found at
https://nowack.lassp.cornell.edu/sites/nowack/files/publications/Tuning%20SC%20CeIrIn5_for_arXiv.PDF

# Figures

## Figure 1: The superconducting transition in a slab under biaxial strain

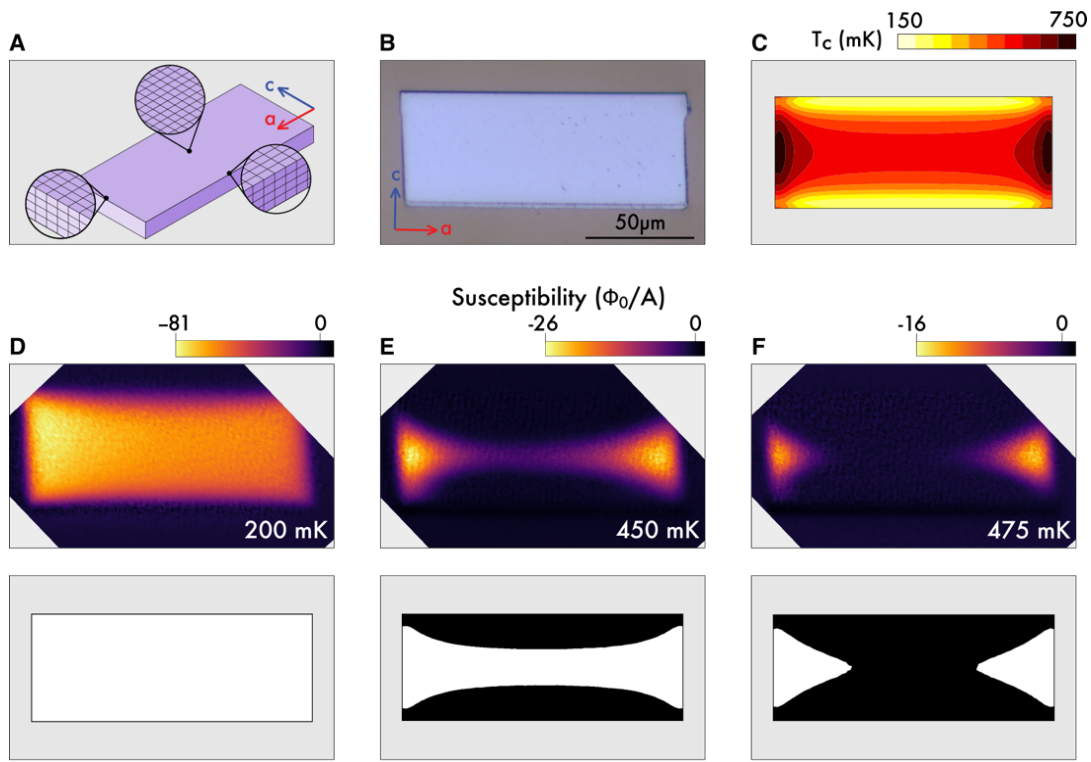

A) Sketch of the distortion of a thin slab of CeIrIn$_5$ coupled to sapphire at low temperatures.
B) Optical image of a 2 μm thick slab cut by FIB machining in the (a,c) plane.
C) T$_c$-map across the sample arising from the strain profile and the strain-dependence of T$_c$ estimated from finite element simulations.
D) – F) top row: Local susceptibility images taken at three representative temperatures. A negative diamagnetic susceptibility indicates superconducting regions of the sample. bottom row: Superconducting regions (shown in white) calculated from the strain profile in the device. The calculated regions correspond to constant temperature contours of the T$_c$-map shown in (C).

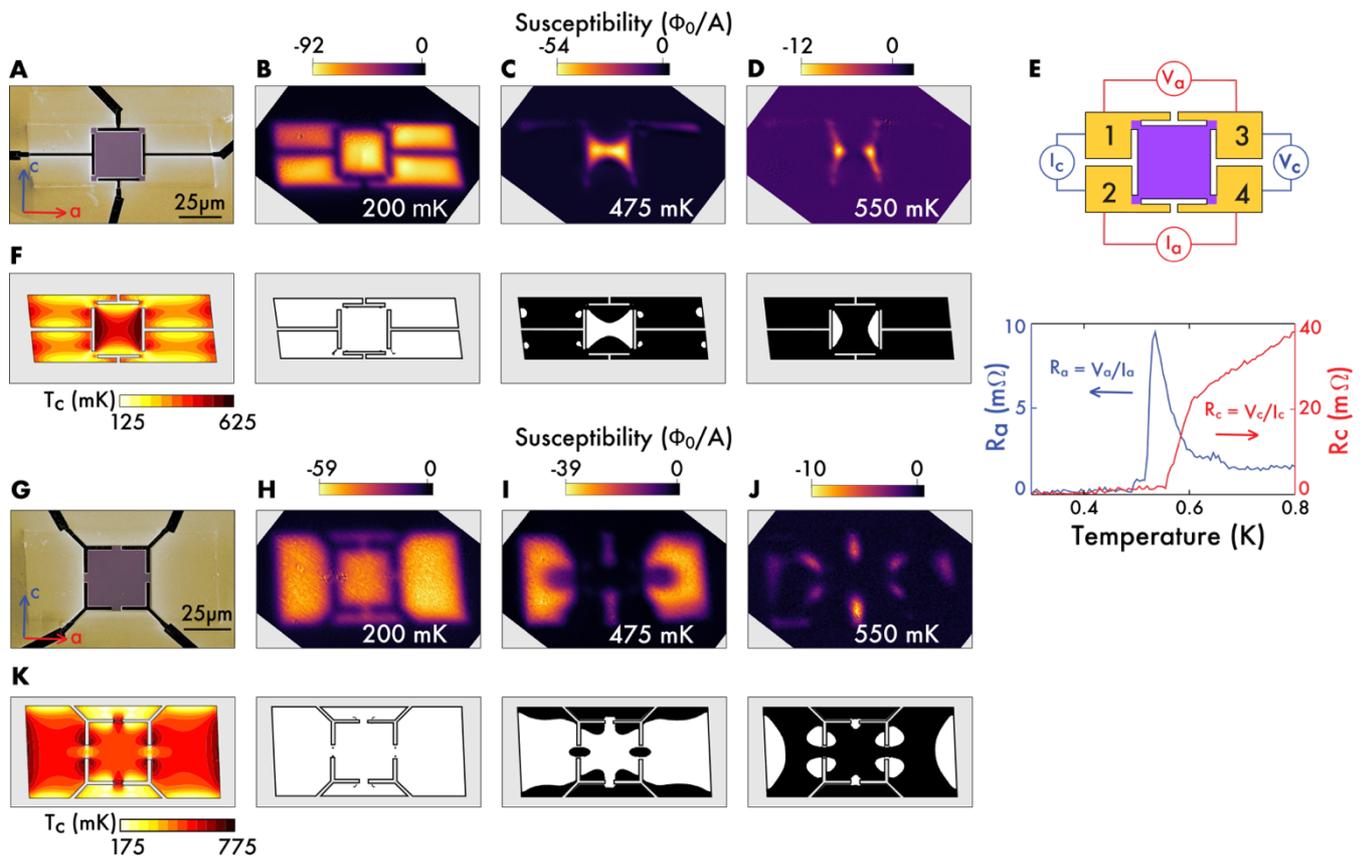

**Figure 2: Spatial control over correlations**

A) SEM-image of device 1. The slab is FIB-cut in the (a,c) plane and contacted by evaporated Au (yellow). The center square of the device is 25x25x3 µm³, held by contacts in the corners. The scale bar is 25µm.

B) – C) Local susceptibility images at three representative temperatures illustrate the temperature evolution of the spatially modulated superconducting state. Lower panels show calculated superconducting patterns.

E) Montgomery transport measurement upon cooling of device 1. As the first superconducting regions appear along the sides along the c-direction (panel D), the c-direction resistance $R_c = \frac{V_{12}}{I_{34}}$ vanishes and the a-direction resistance $R_a = \frac{V_{13}}{I_{24}}$ experiences a voltage spike, owing to a sudden current redistribution. At lower temperatures, these regions touch (panel C), leading to zero resistance across all contacts

F) $T_c$-map from finite element calculations.

G) SEM-image of device 2. It was fabricated as similar as possible to device 1, but with the constrictions connecting at the middle of each side of the square, not the corners (compare panel A).

H) - K) The same as in A)-F), but for device 2. A completely different superconducting pattern unfolds, as expected from the difference in position of the contacts.

**Figure 3: Smaller Structures**

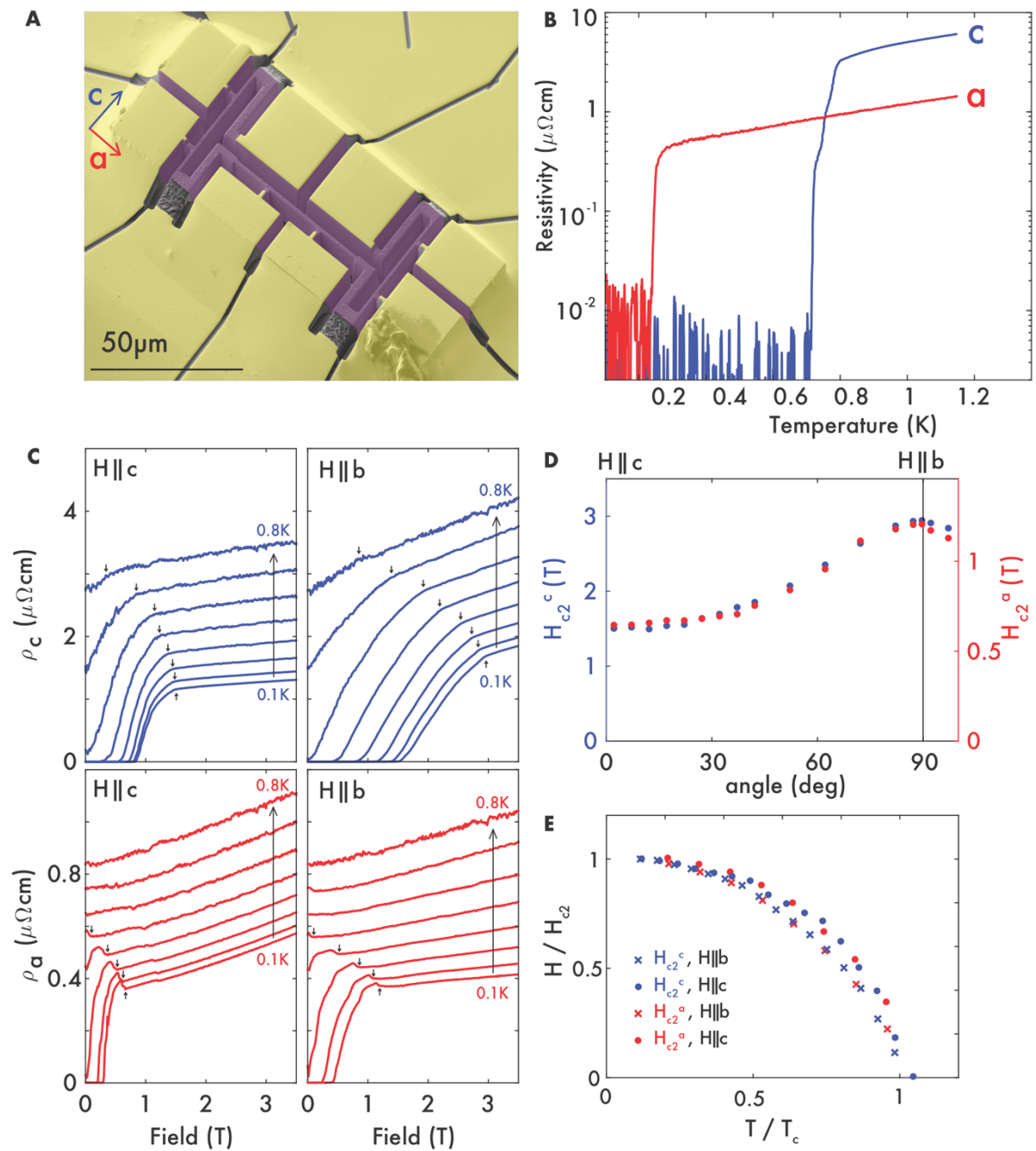

(A) SEM image of device 3. The device consists of long bars of dimension (1.8x8x22 µm³). Two bars are oriented along the c-direction and one along the a-direction.

(B) Resistivity as a function of temperature for device 3.

(C) Temperature-dependent magnetoresistance along the a- and c-direction in $CeIrIn_5$ for fields applied along the c- and b-direction. All traces of $r_a$ were taken in a transverse field configuration, and the b-direction denotes the symmetry equivalent in-plane direction orthogonal to a, the direction of current flow.

(D) Angle dependence of the critical fields defined as the points where deviation from the normal magnetoresistance occurs (arrows in A), for both current configurations. The data collapse onto a single curve by scaling (compare the two vertical axes).

(E) Temperature dependence of the critical fields for all 4 current and field configurations. The curves for each field configuration collapse when scaled by the respective values of $H_{c2}(0K)$ and $T_c$ using the experimental values for in-plane transport ($T_c \sim 0.45K$, $H_{c2}^a||b(0K) \sim 1.2T$, $H_{c2}^a||c(0K) \sim 0.65T$) and along the c-direction ($T_c^* \sim 0.8K$, $H_{c2}^{*,c}||b(0K) \sim 2.95T$, $H_{c2}^{*,c}||c(0K) \sim 1.5T$).

# Figure 4: Robust superconductivity coexists with quantum oscillations

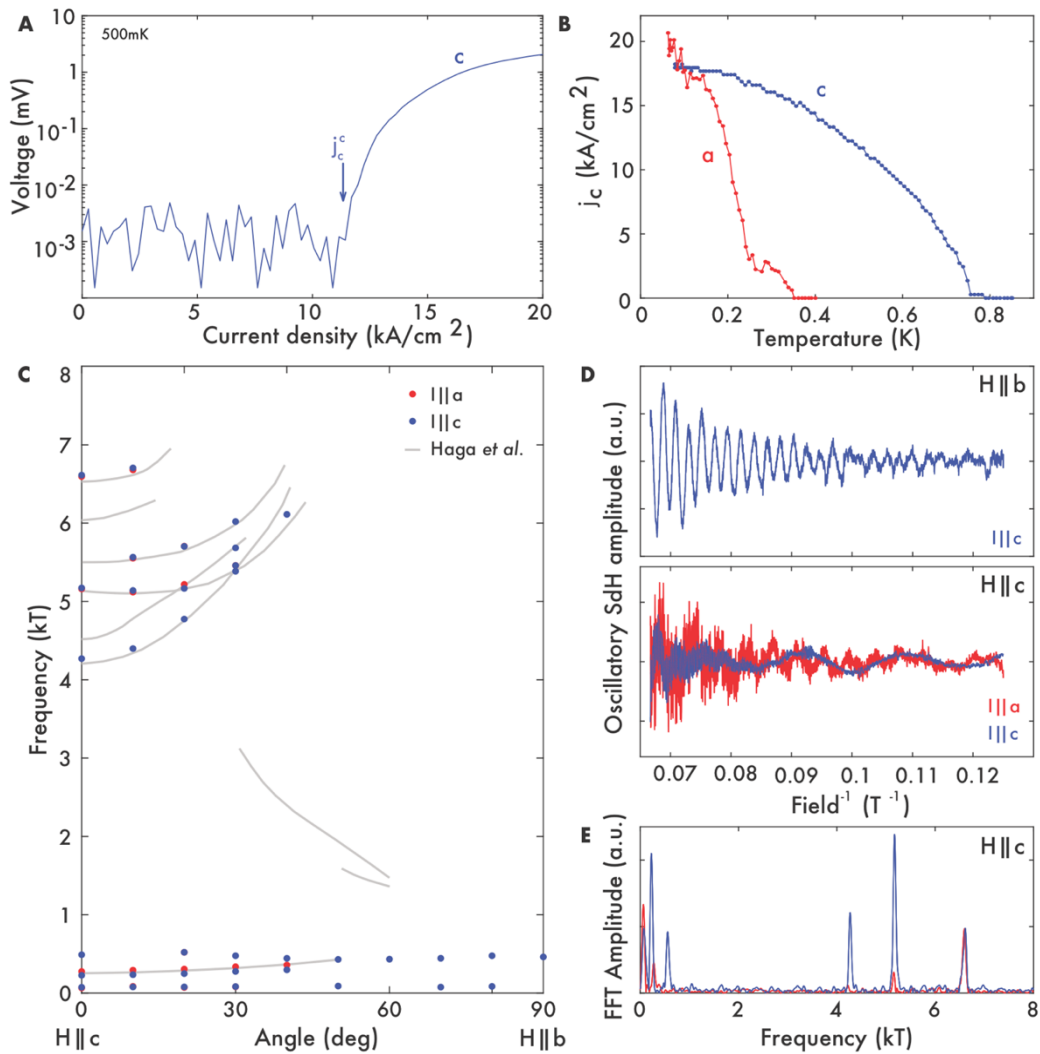

(A) Current/Voltage characteristic along the c-direction measured on the device shown in Fig1A. The onset of measurable voltage above the experimental noise level was used to define $j_c^a$ and $j_c^c$ as indicated by the arrow.

(B) Critical current along both directions. A robust, high-$j_c$ state is observed along the c-direction whereas the a-direction is in a metallic state.

(C) Angle dependence of Shubnikov-de Haas oscillations at 80 mK measured in device 3 (points) overlaid on de Haas-van Alphen oscillations measured on bulk single crystals (grey lines).

(D) SdH oscillation in the microstructure for H∥b (top) and H∥c (bottom).

(E) The FFT spectrum for the oscillations in (D).